# Visualizing Cosmological Concepts
## Using the Analog of a Hot Liquid

E. Yusofi[1] and M. Mohsenzadeh[2]

[1] *Department of Physics, Islamic Azad University-Ayatollah Amoli Branch, Iran*
*P.O.Box 678, Amol, Mazandaran, Iran*

[2] *Department of Physics, Islamic Azad University, Qom Branch, Qom, Iran*

ABSTRACT

We have used the expansion process of hot milk, which has similarities with the cosmic expansion, to facilitate easier and better visualization and teaching of cosmological concepts. Observations of the milk are used to illustrate phenomena related to the Planck era, the standard hot big bang model, cosmic inflation, problems with the formation of structure, and other subjects. This innovative and easily implemented demonstration can enhance the learning of cosmological concepts.

## 1. INTRODUCTION AND MOTIVATION

In texts and monographs of cosmology [1-5] analogies are often used to describe abstract concepts. For example, for the description of the co-moving expansion and the concept of the scale factor often makes use of an inflating balloon, on which points are moving away from each other. Also, cosmologists adopt numerical simulations as effective tools to describe and understand the evolution within the framework of the expanding space-time of the universe. There are various numerical methods − grid or particle based − for performing such simulations. For a recent review of such numerical simulation methods within the cosmological context, see Dolag et al (2008) [6- 8]. In computer simulations the resulting dataset is huge and complex and requires suitable and effective tools to be inspected and explored.

The milk example, by contrast, represents the most immediate and intuitive way to learn and teach cosmology concepts. It offers a simple and inexpensive way to describe early cosmic events. This is not a substitute for systematic analysis, but it accelerates and simplifies the visualizing process. Furthermore our observations (in the milk example) embody several basic elements of cosmology, and so are effective instruments to introduce non-technical people to complicated of concepts.

---

[1] E.yusofi@iauamol.ac.ir
[2] mohsenzadeh@qom-iau.ac.ir

Our prime motivation for introducing the example of milk was the fact that when a filled container of milk reaches the boiling point, it suddenly overflows and therefore has similarities with cosmic inflation. By further study of this natural process, including film and images, we have achieved additional agreements with cosmology. In this paper, we begin with a quick review of the big bang cosmology, and the problems with the standard model, which led to the introduction of inflation [9, 10]. The mechanism of the inflationary epoch using scalar fields is described [11], and then results giving the form of perturbations produced by inflation are quoted. In section 3 we study the overflow process of the hot milk and discuss its super-heating state [12- 14]. In the next section, we compare phases of the cosmic expansion with the expansion of the milk. Finally, we present a list of the agreements, draw conclusions and discuss future work.

## 2. QUICK REVIEW OF COSMOLOGY

2.1. Cosmology

The physical theory used to describe the large scale structure of the Universe and its evolution is the General Theory of Relativity. In deriving a model of the Universe we make use of an assumption called the Cosmological Principle, which says that, "at any given cosmological time, the Universe is homogeneous and isotropic." Homogeneity implies that the geometrical properties of the local neighborhood of any point in space-time will look similar, whereas isotropy means that the universe will look similar in all directions as seen from any point.

*2.1.1. Big Bang Cosmology*

The early Universe begins with the event called the Big Bang. At that time the universe is very hot and dense and as it expands it becomes cooler and less dense. As the Universe starts cooling, it expands and there is a competition between two processes: the deceleration cause by matter and the initial rate of cosmic expansion. The present model of our Universe is based on three major observational facts:

- The uniform Hubble expansion of the galaxies
- The cosmic microwave background radiation
- The observed abundance of light elements

*2.1.2. Cosmic Microwave Background (CMB) Radiation*

The wavelengths of the thermal photons that emerged from the Big Bang are stretched to be in the microwave range owing to the expansion of the universe. This radiation is called cosmic microwave background radiation. We observe it to be isotropic in every direction of the sky today. However, its temperature has been found to have anisotropy of the order of $\Delta T/T \sim 10^{-5}$, imprinted from the initial conditions and serving as the seeds for subsequent galaxy formation.

2.2. Inflationary Universe

Despite its remarkable achievements and widespread acceptance, the Standard Big Bang model of the Universe is not able to give a self-consistent picture of the evolution of our world from the Planck era to the present day period. Some problems have been pointed out during the years and to solve them a new physical theory--inflationary cosmology-- was proposed in 1981 [9].

*2.2.1. Mechanism of Inflation*

According to Figure 1;

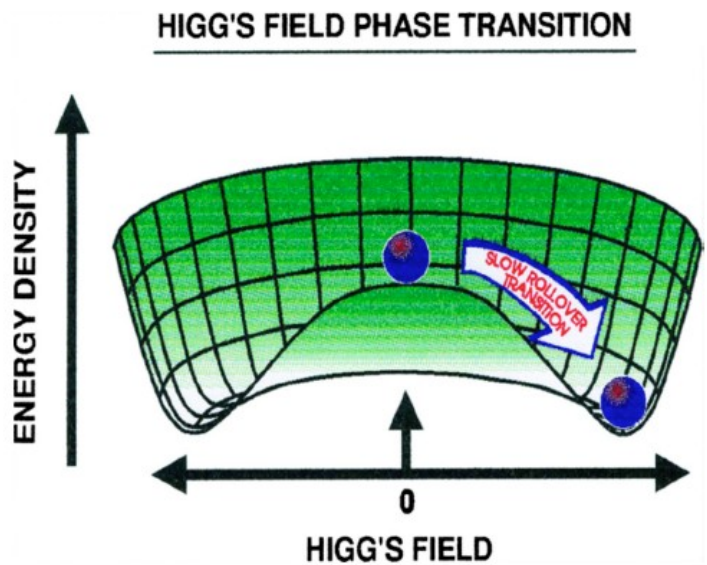

**Figure1:** Mechanism of Inflation

1. In the ordinary vacuum, the Higgs field is non-zero. This is lowest (degenerate) energy state or true vacuum.

2. State $\varphi = 0$ is a meta-stable state and the shape of the Higgs field has unique characteristics. The phase transition is slow compared to cooling rate of the Universe.

3. Regions of the Universe super-cool without breaking symmetry, just as water can super-cool below 273K without turning to ice. The super-cooled regions are in a state called false vacuum.

4. The Higgs field finally reaches its lowest state, the symmetry breaks, and domains of true vacuum eat into false vacuum.

5. True vacuum represents the lowest energy density state (pressure = 0). For true vacuum to expand into false vacuum, the pressure of false vacuum must be negative (i.e., a repulsive force). In this state false vacuum acts like a cosmological constant, as originally described by Einstein and DeSitter. After symmetry breaking, the latent heat stored in Higgs field is released and re-heats the Universe, and finally inflation ends.

2.2.2. *Reheating*

After the end of inflation, the vacuum energy of the field driving inflation was transferred to ordinary particles, so a reheating of the Universe took place. During inflation the Universe itself super cooled exponentially, i.e. $T \sim e^{-Ht}$. The period of inflation and reheating is strongly non-adiabatic, since there is an enormous generation of entropy at reheating. After the end of inflation, the Universe restarts in an adiabatic phase, with the standard conservation of energy. In fact the Universe restarts from very special initial conditions in such a way that the horizon, flatness and monopole problems are avoided.

2.3. Standard Big Bang Problems and the Solution within the Inflation Paradigm

2.3.1. *Flatness problem*

Space-time in general relativity is dynamical, curving in response to matter and energy in the universe. Why then is the universe so closely approximated by Euclidean space?

During the inflationary period, space-time expanded to such an extent that its curvature would have been greatly reduced. Thus, it is believed that inflation drove the universe to a very nearly spatially flat state, which corresponds almost exactly to the critical density.

2.3.2. *Horizon problem*

The cosmological horizon is the farthest distance from which light can have traveled since the Big Bang (i. e., the farthest the observer can observe). We will show shortly that the co- moving Hubble radius $(aH)^{-1}$, characterizes the fraction of co-moving space in causal contact. During the Big Bang expansion $(aH)^{-1}$ grows monotonically and the fraction of the universe in causal contact increases with time. But the near-homogeneity of the CMB tells us that the universe was extremely homogeneous at the time of last scattering on a scale encompassing many regions that a priori are causally independent. How is this possible?

During inflation, the universe undergoes exponential expansion, and the particle horizon expands much more rapidly than previously assumed, so that regions presently on opposite sides of the observable universe are well inside each other's particle horizon. The observed isotropy of the CMB then follows from the fact that this larger region was in causal contact before the beginning of inflation.

2.3.3. *Monopole problem*

The magnetic monopole problem was raised in the late 1970s. Grand unification theories predicted topological defects in space that would manifest themselves as magnetic monopoles. These objects would be produced efficiently in the hot early universe, resulting in a density much higher than is consistent with observations, given that searches have never found any monopoles.

During inflation, the number density of such objects is diluted by the inflationary expansion. The same is true for cosmic strings, domain walls, and all other topological defects.

2.4. Structure Formations

After inflation, space-time expands more slowly than the particle horizon. Thus perturbations will re-inter the horizon when their wavelengths become smaller than the horizon size. The perturbations re-entered the horizon at the matter-dominated period. Since that time they have grown by gravitational attraction and caused structure formation. These original perturbations in the CMB radiation are called *primordial perturbations*.

For convenience, we consider the perturbations on small scales and large scales separately.

2.4.1. *Small scale perturbations*

Small scale perturbations are the perturbations whose wavelengths are smaller than the horizon size at the decoupling time. This means that after crossing outside the horizon during inflation, they have re-entered the horizon before the decoupling time. Therefore the perturbations have been changed because of gravity.

2.4.2. *Large scale perturbations*

Large scale perturbations are the perturbations whose wavelengths are larger than the horizon size at the decoupling time. They have not re-entered the horizon yet, and therefore their amplitudes are the same as in the inflation period.

## 3. THE MILK OVERFLOWING PROCESS

3.1. Why Does Milk Overflow Suddenly?

If you heat milk but don't pay attention to it, it will likely overflow. But you may not understand cause of the overflow. When you heat milk gradually, nothing occurs up to the threshold of the boiling point; but the problem begins with the release of the first tiny bubbles containing the steaming milk. As the temperature rises, these bubbles will form at greater and greater speed, and finally when the speed of the formation of these bubbles is more than their expansion speed, the volume occupied by the bubbles increases rapidly and after a time they pour out from the container. Bubbles are also made in water, but they pop when they reach the surface of the water, releasing invisible steam. The surface tension of milk is more than water because of the existence of proteins, fats, and carbohydrate compounds. This higher surface tension causes more resistance of bubbles against the internal steam pressure and increases their lifetimes.

3.2. Milk or Water That Doesn't Inflate

If you heat a smooth container of milk or water in a microwave oven, it won't overflow although its temperature gets higher than 100° C. A microwave oven produces waves at a special frequency, which causes the periodic movement of water molecules in milk, making friction and causing homogeneous heating (i.e. at any time, all parts of the liquid are isothermal in the microwave oven.). Thus, water or milk in a glazed container in the microwave oven starts heating quietly. This way of heating can continue to the boiling point without the formation of any bubbles.

Water boils at 100 °C if there is already a bubble of steam (or air) present. But in the absence of bubbles, water can be heated above 100 °C. This state is called *super-heating*, and it isn't special to the microwave oven but it results from a deeper physical fact. If the water is completely homogenous, there won't be any bubbles in the water, and it continues to get hotter without boiling. However, if the symmetry of this water is disturbed in any way (for example, when objects (e.g. a spoon) or granulated materials (e.g. instant coffee) are put into it, the water may boil very vigorously or even appear to explode out suddenly of the container.

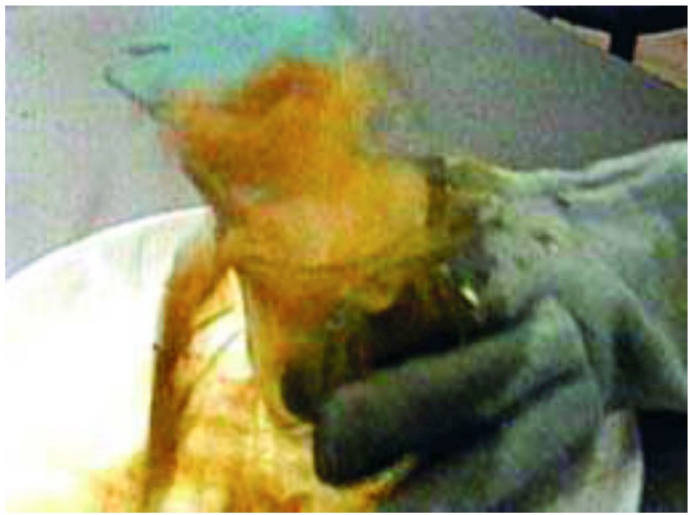

**Figure 2:** Super heating water

There is a common belief that superheating can occur only in pure substances. This is untrue, as superheating has been observed in coffee and other impure liquids. An impurity such as salt or sugar, dissolved in water to form a homogeneous solution, will not prevent superheating [14].

## 4. THE EXPANSION OF THE HOT UNIVERSE COMPARED TO THE EXPANSION OF HOT MILK

Consider the milk container on the flame of the stove. After a time, first milk starts boiling slowly and produces tiny bubbles. Then these bubbles inflate and merge, and the volume of the milk in the container increases. If the height of the container is low and the amount of the fat heterogeneous, the milk will overflow (Figure 3). Some bubbles will remain in the container, and if heating continues, the milk will continue to overflow.

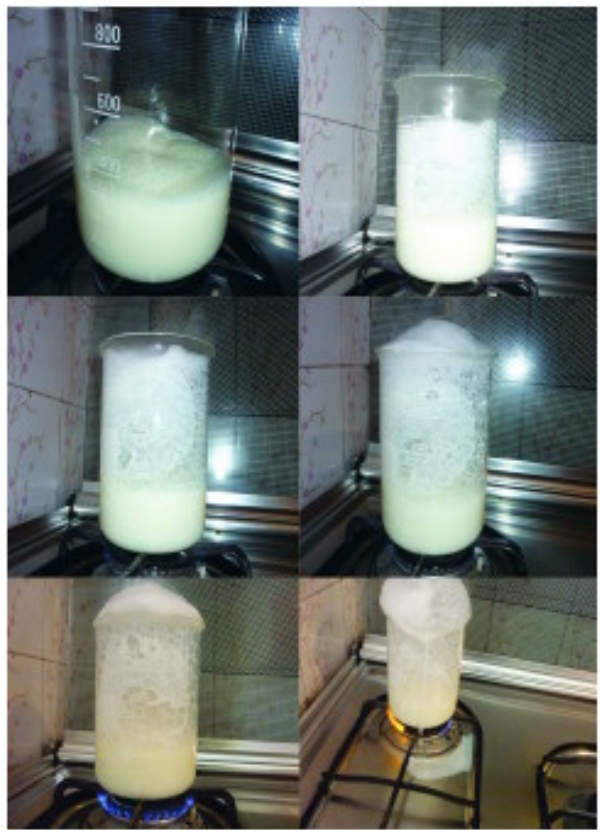

**Figure 3:** The heating of milk and its expansion

## 4.1. Expansion Eras of the Milk and Universe

### 4.1.1. *The Planck Era*

At the Planck time $t \sim 10^{-43}$ (s), the Universe has space-time fluctuations, which must be investigated within the framework of quantum gravity theory. This phase is analogous to the time when tiny bubbles are forming in the milk. Both milk and its surface are are starting to have fluctuations but the milk is yet inflating.

### 4.1.2. *The Inflationary Era:*
The time period is $10^{-43} < t < 10^{-35}$(s). In a tiny fraction of a second, the small regions of space-time fluctuations begin to expand exponentially. In the milk example, in a short time the bubbles merge and expand suddenly and encompass the major part of the volume of the container and even may pour out of it.

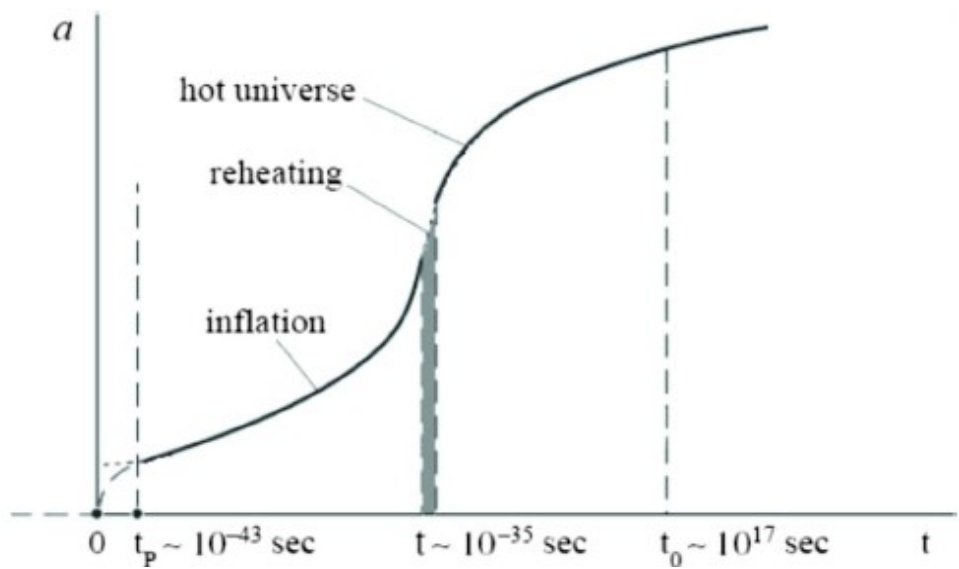

**Figure 4:** Expansion eras of the universe (alpha is a dimensionless scale factor)

### 4.1.3. *Reheating*
According to Figure5,

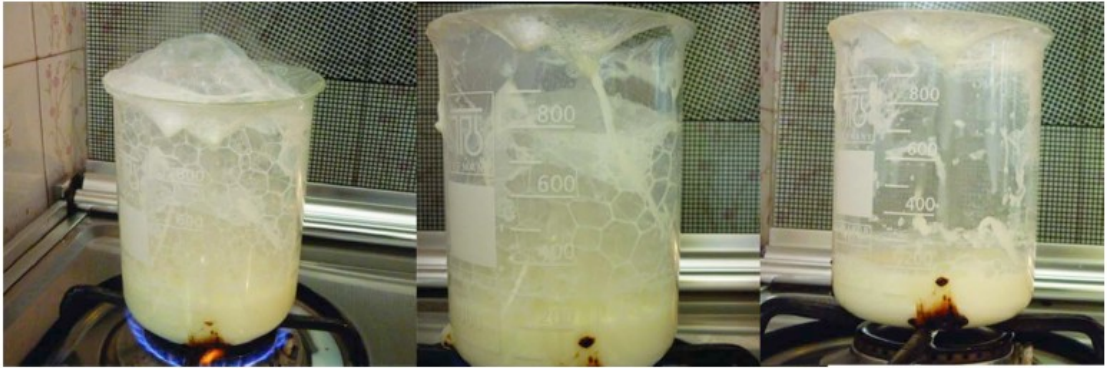

**Figure 5:** Reheating in milk example

After the end of inflation of milk, some of the bubbles pour out from the container and the remaining bubbles in the pan are in the new ground state. If the heating continues, milk again will do another other phase transition, but this time the expansion is less powerful because milk gets more diluted after the first overflow. We note that in the milk example, this occurs because of loss of material, which is not true in the Universe.

4.2. *The Inflation Mechanism in the Milk Example*

We consider the super heating state as analogous to the false vacuum, therefore:

1. When the bubbles form, the walls of bubbles have a nonzero amount of energy due to surface tension but before the forming of the bubbles, there is zero energy in the ground state.
2. When the milk is superheated but before it inflates, there are no bubbles and all of the liquid is homogenous in terms of the density and there is no preferred point in the milk or on its surface.
3. The inflation of the milk is originally caused by the escape of energy from the heating source (super-heating state) into a medium at the ground state with a lesser energy. The phase transition operates slowly but the whole time of this energy transmission in comparison with the time spent in boiling the milk is small.
4. We prefer the phrase "the regions of the super-heated Universe occur without any breaking of the symmetry", since in the Planck era the universe was in the super-heated condition because of its very high temperature and density. This super-heated milk remains quiescent and has the perfect symmetry and homogeneity, but it over flows explosively with the smallest stimulation and inhomogeneity.
5. In the big bang, matter fields get trapped in a false vacuum state from which they can only escape by nucleating bubbles of the new phase, or the true vacuum state. In the universe, bubbles of the new phase expand and coalesce until they take over the medium and the phase transition is complete. In the milk, the bubbles of steam rapidly expand and combine to increase their volume by a large factor and drive an overall expansion.
6. Because liquid does have the equivalent of an unstable vacuum in the super heated or super-cooled state, it may expand until its temperature falls to the boiling point and its true ground state. This means that the expansion of the hot milk is also an analogy for the cooling of the universe as it expands.

## 5. CONCLUSIONS

Visualizing is one of the most effective techniques to explore and present scientific data, understanding at a glance its basic features and properties. In this paper, we have shown how the milk example can be compared to some concepts of early universe cosmology. This example has the ambitious goal of describing the birth and early evolution of the Universe through the use of an effective and easy to understand simulation of cosmological concepts. Analogies like this help to give the general public or students studying cosmology insight in our model of the Universe on its largest scales as derived from its earliest physics.

ACKNOWLEDGMENT

We are indebted to Prof. Sidney Wolff and Prof. Chris Impey for helpful conversations and new suggestions in this version of paper. Also, we would like to thank M. V. Takook, J. Sadeghi and F. Motevalli for their interest in this work.